# Understanding defects in amorphous silicon with million-atom simulations and machine learning

Joe D. Morrow,[1] Chinonso Ugwumadu,[2] David A. Drabold,[2] Stephen R. Elliott,[3] Andrew L. Goodwin[1,]* & Volker L. Deringer[1,]*

[1] *Inorganic Chemistry Laboratory, Department of Chemistry, University of Oxford, Oxford OX1 3QR, UK*

[2] *Department of Physics and Astronomy, Nanoscale and Quantum Phenomena Institute (NQPI), Ohio University, Athens, Ohio 45701, United States*

[3] *Physical and Theoretical Chemistry Laboratory, Department of Chemistry, University of Oxford, Oxford OX1 3QZ, UK*

\* andrew.goodwin@chem.ox.ac.uk; volker.deringer@chem.ox.ac.uk

**The structure of amorphous silicon is widely thought of as a fourfold-connected random network, and yet it is defective atoms, with fewer or more than four bonds, that make it particularly interesting. Despite many attempts to explain such 'dangling-bond' and 'floating-bond' defects, respectively, a unified understanding is still missing. Here, we show that atomistic machine-learning methods can reveal the complex structural and energetic landscape of defects in amorphous silicon. We study an ultra-large-scale, quantum-accurate structural model containing a million atoms, and more than ten thousand defects, allowing reliable defect-related statistics to be obtained. We combine structural descriptors and machine-learned local atomic energies to develop a universal classification of the different types of defects in amorphous silicon. The results suggest a revision of the established floating-bond model by showing that fivefold-coordinated atoms in amorphous silicon exhibit a wide range of local environments, and it is shown**

**that fivefold (but not threefold) coordination defects tend to cluster together. Our study provides new insights into one of the most widely studied amorphous solids, and has general implications for modelling and understanding defects in disordered materials beyond silicon alone.**

Amorphous silicon (a-Si) is the textbook example of a disordered material, with a structure that overall resembles Zachariasen's concept of a continuous random network (CRN) of covalently bonded atoms.[1–3] Many interesting physical properties and phenomena relating to a-Si have been discussed over the years: the hyperuniform nature of the disordered network,[4] the transition between the high-coordinated metallic liquid and the fourfold-connected structure of a-Si,[5] its complex phase behaviour under gigapascal pressures,[6–8] and a tension–compression asymmetry in its mechanical properties.[9] Many of those phenomena originate on the atomic scale and can be rationalised in terms of local coordination environments, point-defect formation, and chemical bonding. Fully understanding the atomic-scale interactions in disordered solids, particularly including a-Si, and revealing how they determine macroscopic material behaviour has long been a central research goal.

While the overall structure of a-Si is based on fourfold-connected atoms ($N = 4$, where $N$ denotes the number of bonded nearest neighbours), there is particular interest in those 'defective' atoms that have fewer ($N = 3$) or more ($N = 5$) neighbours in the first shell. In well-relaxed structural models of a-Si, $N$ is easily determined: the structure shows a clear minimum in the radial distribution function, separating the first (bonded) peak from the second (non-bonded) one, at a distance of about 2.85 Å. The counting of neighbours and therefore the definition of coordination defects is straightforward if the minimum in the distribution decreases to zero; it becomes more ambiguous otherwise, which is most relevant to $N = 5$ environments. There is also much experimental evidence—for example, from spectroscopy— for the presence of point defects in this amorphous material, emphasising that the structure of a-Si is more nuanced than a simplified all-fourfold CRN description.[10–15]

Computer simulations have long played a key role in studying amorphous networks, and a-Si has been a prominent example.[16] Back in the 1980s, structural models of a-Si were created *via* melt-quench molecular dynamics (MD) simulations with the empirical Stillinger–Weber potential.[17,18] The large concentration of coordination defects in the models obtained this way ($\approx 20\%$) permitted a discussion of the average structure of $N = 5$ defects in terms of bond angles, as well as an analysis of the energetics predicted by the potential. Later studies, with increasingly fast computers, extended the system size to hundreds of thousands of atoms with empirical potentials.[19–21] In a complementary vein, quantum-mechanically (density-functional theory, DFT) based MD studies on much smaller systems have also provided important insights.[22,23,24] However, beyond the few-nanometre length scale, predictive DFT simulations of a-Si remain out of reach, due to the long simulation times required to capture the slow dynamics in silicon, and to the unfavourable cubic scaling of the computational cost with system size. For example, a current 'optimal' DFT-MD-based a-Si structure contains 215 atoms,[25] and the current limit for such types of MD simulations in general appears to be about 1,000 atoms even on fast supercomputers.[26]

Recent developments in machine-learning (ML) based interatomic potentials have made it possible to prepare realistic, DFT-accurate structural models of a-Si of much larger size,[8,27–30] with a recently published million-atom model reaching a cell length of about 27 nm.[29] We have previously established that slow quenching from the simulated melt using ML potentials yields structural models whose characteristics are in good agreement with existing experimental data.[8,27,29] There is also an increasing body of evidence that the local, per-atom, energy predictions obtainable from atomistic ML methods (but not normally from DFT) are amenable to post-hoc chemical interpretation: we have shown that these ML atomic energies can be used to discriminate three- and five-coordinated atoms in a-Si in an initial pilot study,[28] that they can be used to define a local enthalpy,[8] and that they can also be used to drive structural exploration.[31] Neural-network models have been interrogated with regards to local energy contributions as well.[32–34] A recent study has shown that the density of ML atomic-energy contributions can yield insights into complex solid-state ion conductors.[35]

In the present work, we report a major step forward in understanding the centrally important structural point defects in a-Si, with implications for disordered materials more generally. ML-driven simulations allow us to overcome a fundamental problem in studying defects with small-size simulations: namely, that each individual simulation cell is subject to large statistical variation. Having comprehensive statistics available through million-atom simulations, we now identify three structural prototypes for over-coordinated atoms in a-Si, and we explain the tendency for such defects to aggregate via the strain that they induce on their atomic neighbours. Our study paves the way for routine quantum-accurate, million-atom scale, ML-driven simulations of rare events such as defect formation in functional materials.

## Results

### Quantum-accurate million-atom models

Our simulations of a-Si are based on a computational approach which we call 'indirect learning' (Fig. 1a),[29] and which corresponds to teacher–student models for knowledge distillation that are more commonly used in ML research. The simulation approach has been validated in our previous, more technical work,[29,36] and we use it here to set the stage for an in-depth analysis of the defects in a-Si. Indirect learning involves using an accurate, trusted, but computationally slow ML potential (the teacher model) to train a second, much faster one (the student model). The latter enables simulations of accuracy on a par with that of the teacher, whilst requiring much less computational time, by a factor of about 1,000 in this case. A visual comparison between 100,000-atom and 1M-atom models, drawn to scale in Fig. 1b, illustrates the advantage of the approach. Million-atom simulation cells are necessary to systematically study defects that occur at the few-percent level, and, crucially, to investigate 'second-order' processes, such as the interaction of defects with one another. Here, the simulation cell contains tens of thousands of examples of threefold- and fivefold connected defective atoms (Fig. 1c).

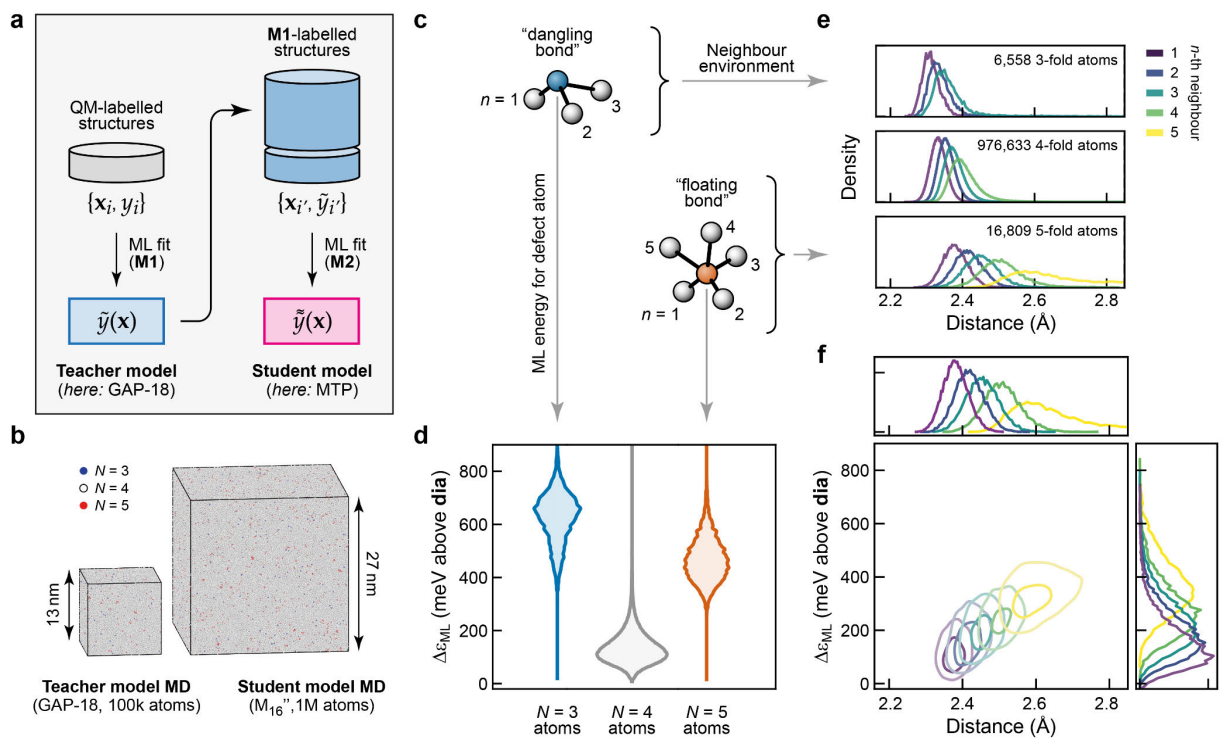

**Fig. 1: Defects in amorphous silicon from million-atom simulations.** (**a**) Schematic of the teacher–student approach to ML potential fitting described in ref. [29]. A reliable, but relatively slow teacher model (**M1**) is used to generate a set of many more structures, to which the more specialised, but faster student model (**M2**) is then fitted. (**b**) Visualisation of 100,000- (*left*) and million-atom (*right*) structural models of a-Si, drawn to scale for direct comparison. Panels (a) and (b) are adapted from ref. [29], which was originally published under a CC BY licence (https://creativecommons.org/licenses/by/4.0/). (**c**) Schematic drawing of 3- and 5-fold-connected atoms, with neighbouring atoms numbered in the order of their distance from the central atom. (**d**) Distributions of the ML-predicted atomic energies, $\varepsilon_{ML}$, of defects in the million-atom model, shown separately for different nearest-neighbour coordination numbers, *N*. All values are referenced to crystalline diamond-type silicon, which is set as the energy zero. (**e**) Neighbour distributions, resolved according to 3-, 4-, and 5-fold-connected central atoms (separate axes) and their individual immediate neighbours, sorted by distance. (**f**) 2D correlation plot of the neighbour density for 5-coordinated defects (as in panel e) versus energy (as in panel d), given separately for the immediate neighbours ($n$ = 1–5, purple to yellow).

Our a-Si structural models show quantitative agreement with available structural experimental data: the inverse height of the first sharp diffraction peak is $H^{-1} = 0.564$ for the 100,000-atom structure from ref. [8], and $H^{-1} = 0.560$ for the million-atom model from ref. [29], compared to experimental data of $H^{-1} = 0.565$ (ref. [4]). We carried out electronic-structure analyses for additional validation, including a million-atom tight-binding computation, as discussed in the Methods section.

Perhaps the most serious assumption made with most current ML potentials is that the total energy of a system of atoms (which is a quantum-mechanical observable) can be decomposed into a sum of local atomic energies (which are not observables).[37,38] The major benefit of making this locality assumption is the ability to 'machine-learn' energetics independent of the system size, as well as the resulting linear scaling of computational cost with the number of atoms in the simulation. Although initially conceived for this purpose alone, there is now increasing evidence that local energies have a physical relevance that can be useful for analysis beyond the mere construction of fast ML potentials,[28,35] and that their robustness can be quantified.[39] We note that the idea of considering local contributions to the total energy in silicon has been pioneered based on empirical potentials,[17] and our present ML-based approach extends this type of thinking and places it on a more quantitative, DFT accurate basis.

Figure 1d shows the distribution of ML atomic energies in the 1M-atom model of a-Si from ref. [29], separated according to coordination numbers for the central atom. It confirms earlier findings that $N = 3$ atoms have distinctly higher energies on average than do $N = 5$ environments.[28] Note, however, that the distributions in Fig. 1d result from a much larger dataset than in ref. [28], and therefore do not include any broadening.

Figure 1e presents an analysis of the defects in a-Si from an alternative, and somewhat orthogonal, purely structural perspective, by examining the distances to individual neighbours for each atom in the structure. As the coordination number of the central atom increases from 3 to 5, so too do the distances between bonded atoms, as is reasonable from a chemical perspective. The well-defined 5-th peak, and the even spacing between peaks for the 5-fold distribution, indicate that a majority of such defects are best described as 'truly' 5-coordinated, rather than alternatively as [4+1] with a mostly non-bonded 5-th neighbour as one would expect in tetrahedral amorphous carbon.[40] The longer tail observed for the $n = 5$ peak contains those more distant 5-th neighbours that indeed are in this minor group of [4+1] environments. The typical bond lengths for $N = 3$ and $N = 5$ defects in our a-Si model differ strongly from those for the non-defective 4-fold atoms in the rest of the structure, consistent with the observation that defects have significantly higher energies than do most bulk-like atoms (Fig. 1d). In Fig. 1f, we examine the correlation between those energetic and structural indicators, *viz.* the local energy of the neighbours of the 5-fold atom on the *y*-axis, and the distance of the respective neighbour from the central atom on the *x*-axis. We find a striking, approximately linear correlation between these two quantities, which shows that the elevated local energy at 5-fold defect centres is also delocalised onto the surrounding atoms with longer bonds to the defect. The equivalent correlation plots for $N = 3$ and $N = 4$ atoms (Supplementary Fig. 6) show a much tighter distribution of energies for the respective atomic neighbours.

**Revision of the floating-bond model**

The presence of some relatively short fifth-neighbour contacts evident in Fig. 1e, at distances below 2.6 Å, invites the question whether there is a 'typical' geometric structure of 5-fold defects in a-Si. In fact, this question has been studied for a long time. The term 'floating bonds' has been used to describe defects consisting of silicon atoms with five bonded neighbours, initially introduced by Pantelides in 1986 as part of the first recognition of the importance of over-coordinated atoms in a-Si.[41] The canonical structure of such a defect was described as a perfect tetrahedron with an extra fifth atom bonded directly opposite to one of the equivalent existing bonds. However, it was already noted that, in the amorphous phase, bond distances and angles can vary considerably from such an idealised geometry.

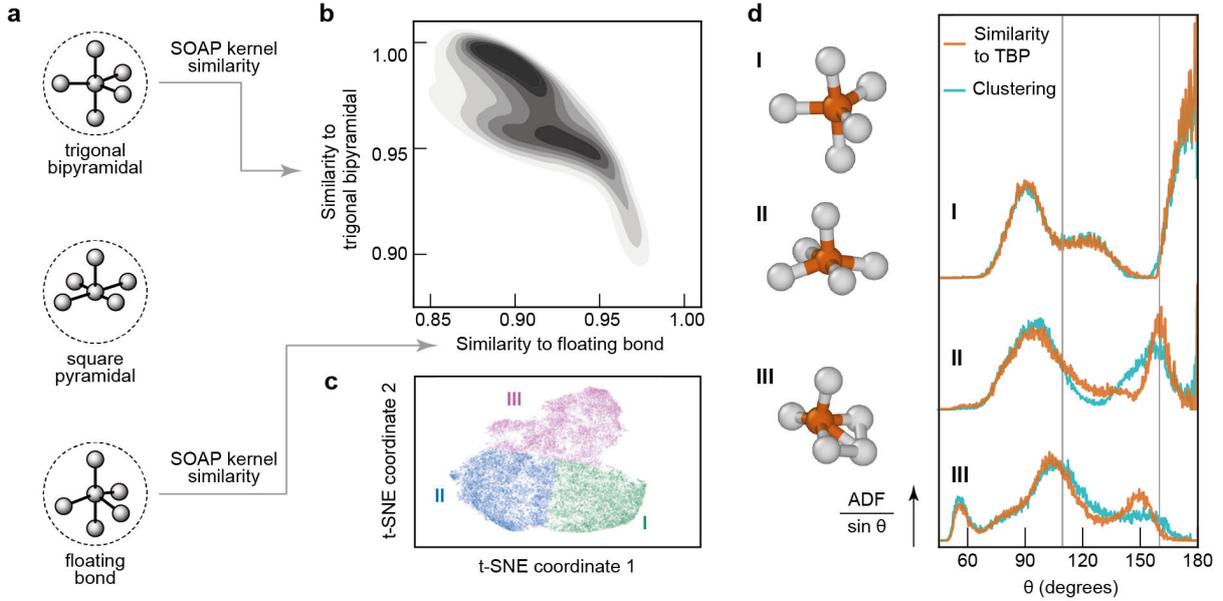

**Fig. 2: Categories of fivefold-connected defects.** (**a**) Schematic drawing of idealised trigonal bipyramidal (TBP), square pyramidal, and 'floating-bond' environments. The first two are shown in line with the established valence shell electron-pair repulsion (VSEPR) model;[42,43] the third is a tetrahedral environment with a fifth atom directly opposite a bond.[41] (**b**) 2D plot of the SOAP kernel similarity of $N = 5$ atoms in the a-Si model to the idealised TBP and floating-bond environments, respectively. The distribution of values for individual atoms is shown as a heat map. (**c**) Unsupervised classification of 5-fold atoms. The full distance matrix, $\mathbf{D} = \sqrt{2 - 2\mathbf{K}}$, is embedded in 2D with the dimensionality reduction algorithm t-SNE,[44] where $\mathbf{K}$ is the kernel matrix built from the similarity of each 5-fold atom with every other 5-fold atom. The bisecting *k*-means algorithm is used to identify clusters I-III.[45] (**d**) Bond-angle distribution function (BADF) plots, scaled by $\sin\theta$, for atomic triples centred on all 5-fold coordinated atoms. The BADFs are plotted separately for the three distinct categories of $N = 5$ defects related to idealised structures respectively from top to bottom (as illustrated with selected examples of such configurations from the a-Si model). The 5-fold atoms are separated into these categories via two methods but with similar results: by comparison to the idealised structures of panels **a** and **b** via SOAP similarity (in orange) and from unsupervised clustering (in cyan). Vertical lines at the tetrahedral angle (109.5°) and at 160° are guides for the eye.

In Fig. 2, we address this question with reference to the common prototypes for 5-coordinated atoms that are known from structural inorganic chemistry. On the left of Fig. 2a, we sketch the trigonal bipyramidal (TBP) and square pyramidal environments that one would expect from the valence shell electron-pair repulsion (VSEPR) model[42,43] that is frequently discussed in undergraduate textbooks, and also an idealised floating-bond environment with a fifth bond directly opposite one of the bonds in a tetrahedral environment (Fig. 2a).[41]

We took 10,000 isolated 5-fold defects in the 1M-atom structural model and evaluated their structural similarity to the three respective prototypes on a scale of 0 (dissimilar) to 1 (identical) *via* the SOAP kernel,[46] as detailed in the Methods section. In this analysis, we aimed to discriminate environments based on their bond angles and their respective similarity to the prototypical VSEPR configurations; we therefore re-scaled all nearest-neighbour distances to 2.5 Å, such that we examine only the angular distribution of atoms around each defect. A multimodal distribution of similarity values results, as displayed in Fig. 2b. By dividing this distribution into three regions, we arrive at a classification of the 5-fold defects in a-Si as TBP-like (**I**), square-pyramidal-like (**II**), or 'floating bond'-like (**III**). A very similar classification is obtained using unsupervised ML[47,48] in Fig. 2c: dimensionality reduction followed by clustering. In this plot, the distance between points corresponds to their structural dissimilarity as measured by SOAP. The three-lobed distribution, with higher densities of points at the outer edges, supports the identification of exactly three major classes of 5-fold atoms.

We show bond-angle distribution functions (BADFs) and accompanying representative example structures taken from the model in Fig. 2c. We note that the BADFs in a-Si have been closely linked to the Raman transverse-optic peak width,[49] to the exponential tails in optical-absorption band edges, and to the through-bond (topological) distance to over- and under-coordinated atoms.[50] Our million-atom simulations allow us to derive finely resolved BADF plots,[29] revealing immediately that the different categories have distinctly different shapes. The BADF for category **I** defects in Fig. 2c directly reflects the angles in an idealised TBP environment: 180° between the axial atoms, 120° between equatorial atoms, and 90° between axial and equatorial atoms. The square-pyramidal BADF (category **II**) has a fairly sharp distribution of angles at 160° and a broader peak at ≈ 100° angles. The designation of a 'floating bond' (**III**), maximally dissimilar from the TBP (**I**), is less well-defined structurally, but can be understood as originating from an ideal tetrahedron with an additional 5-th atom approaching a face or edge, as originally suggested.[41] In the BADF, this arrangement manifests in the following features: (i) the occurrence of ≈ 60° angles between the 5-th atom and those neighbours that it approaches most closely; (ii) a corresponding closing of the ideal tetrahedral angle as these neighbours are displaced away from the 5-th atom; and (iii) larger angles between the 5-th atom and those on the opposite side of the tetrahedron. Fewer than half of all isolated $N = 5$ defects adopt a structure similar to the floating-bond description.

The boundaries between the classes of 5-fold defects, however, are somewhat fuzzy and this suggests an overlap of types of defect structures, rather than three entirely separate categories, as expected for such a strongly disordered amorphous structure. This observation is consistent with the low energy barriers between structural minima that are observed for 5-fold complexes in molecular inorganic chemistry, which often exhibit fluxionality.[51–53] Further discussion of where we draw the boundaries between classes, along with partial pair correlation functions for defects, can be found in Supplementary Fig. 2.

**Locality of defect environments**

The correlation between bond distance and local energy in Fig. 1f had already suggested that the mechanical strain associated with 5-fold defects extends over a longer range than the radius of the defect centre itself. Figure 3 now provides a more detailed analysis of the atomic energies and their degree of 'locality'—that is, their dependence on nearest-, next-nearest-, and further neighbours. The locality of physical properties is of general interest for the development of atomistic ML models, because it determines directly the extent to which information can be 'learned' by finite-range models.[54]

We introduce some notation to describe the topological neighbourhood of defective atoms, as sketched schematically in Fig. 3a: 4-fold coordinated atoms that are directly bonded to a defect centre (*i.e.*, to an under- or over-coordinated atom) are referred to as 4′; 4-coordinate atoms directly bonded to a 4′ atom, but not to a defect centre, are referred to as 4″; those bonded to a 4″ atom, but not to any closer defect, are referred to as 4‴. Atoms which are more topologically distant than the third neighbour shell, 4‴, we merely call 'bulk'.

Figure 3b indicates how, upon moving away from an $N = 3$ defect atom, the atomic-energy distributions quickly approach those of the bulk. The directly adjacent atoms, 4′, have a distribution that is similar to that of the bulk atoms (shown in light grey)—in other words, an $N = 3$ atom does not seem to notably affect the atomic energies of its directly bonded neighbours. In contrast, Fig. 3c shows that the 4′ atoms connected to an $N = 5$ defect are much higher in energy on average. This suggests that the structural disturbance of defects in this case is less localised, and has a longer-range effect. In both cases, the energy distributions for the 4″ atoms and beyond are not notably affected by the presence of nearby defects.

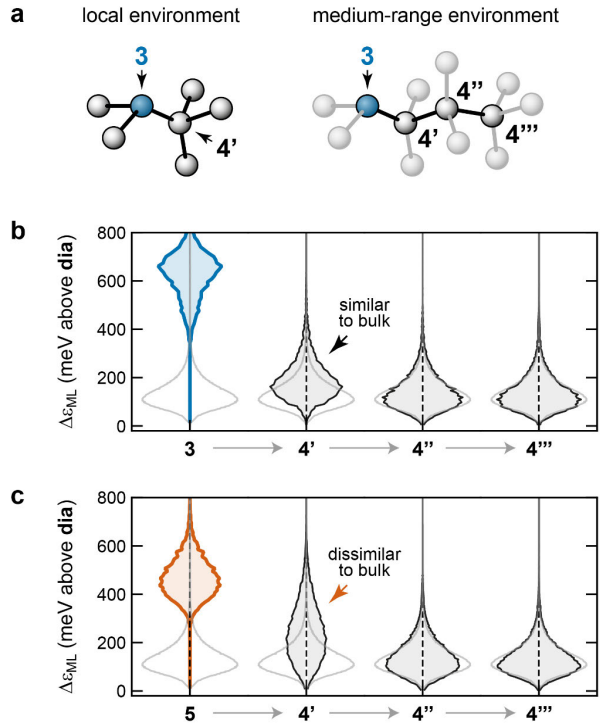

**Fig. 3: Locality of defect environments.** (**a**) Schematic of the labelling scheme for local and medium-range environments, based on the bond topology. A fourfold atom directly connected to a defect (here, to a 3-coordinate atom) is labelled as 4′, a fourfold atom connected to that one is labelled as 4″, and so on. (**b**) Distribution of ML local energies for $N = 3$ atoms and their surroundings. The distribution for bulk a-Si is shown in light grey. (**c**) Same but for $N = 5$ atoms and their surroundings.

## Connection with defects in crystals

The structural and energetic landscapes of defects, even for crystalline materials, can be highly complex, and their exploration and understanding requires advanced computational techniques. For example, it was shown recently that relaxations of small-scale defect models often determine incorrect (metastable) defect geometries, and that a more comprehensive exploration of the associated structural and energetic landscape is required even for seemingly simple inorganic crystals.[55–57] ML techniques have begun to be applied to defects in crystalline materials as well.[58–60] Here, we investigate the question whether there is a connection between defects in crystalline silicon, which are well-studied, and the atomic environments of defects in the amorphous phase.

We have previously shown that 2D scatter plots of structural properties (specifically, the 'diamond-likeness') and atomic energies are useful for understanding defects in a-Si.[28] We now extend this methodology by using it to compare defects in the amorphous form of silicon to different types of defects in the crystalline phase, as shown in Fig. 4. The results are broadly in line with expectations: the 3-fold defects are most similar to vacancies (blue in Fig. 4a), 5-fold defects are most similar to crystalline interstitials (orange in Fig. 4c), and neither defect is very similar to the perfect crystal, as measured by the SOAP-kernel similarity on the horizontal axis. Note that the interstitial example was produced by equilibrating an idealised 10-fold coordinated interstitial site in diamond-type Si at 500 K. These simulations show that the structure of the crystalline interstitial, like the amorphous 5-fold defect, is highly fluxional and is therefore difficult to understand via even several 'typical' structures. In turn, this observation highlights the advantage of having many examples of individual defects in the million-atom model available for comparison, as we have discussed in the context of Fig. 3.

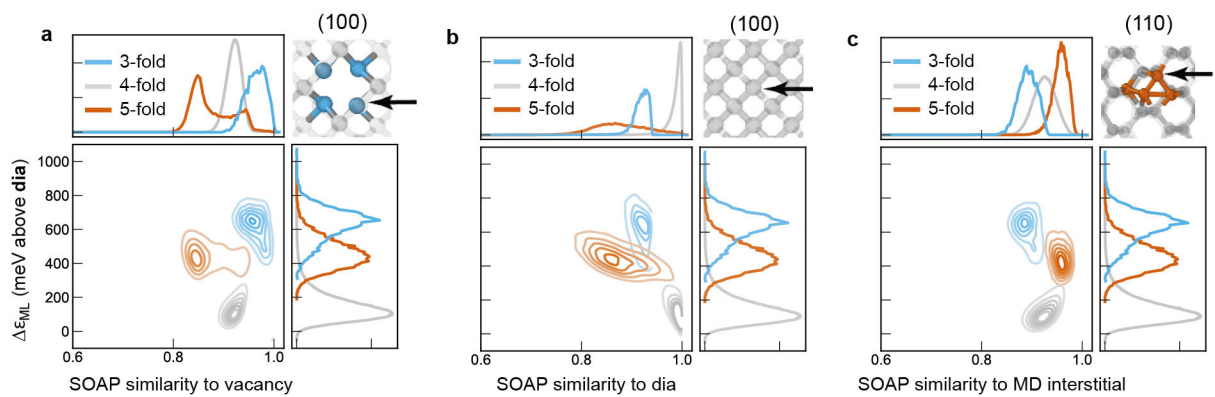

**Fig. 4: Connection with defects in crystals.** (**a**) A 2D SOAP similarity–energy map for the structure obtained at a quench rate of $10^{11}$ K s$^{-1}$, analysing separately 3-fold, 4-fold, and 5-fold coordinated atoms. The horizontal axis shows the structural similarity to the relaxed vacancy; the *N* = 3 atoms (*blue*) are structurally the most similar to the vacancy; the *N* = 5 atoms (*orange*) the least so. The vertical axis gives the ML atomic energy. The structure is visualised in the top right part, with the viewing direction given, and the arrow indicating the atom used for comparison. (**b**) Same for the perfect crystalline diamond structure, as in ref. [28]. (**c**) Same for a snapshot from an MD simulation of an interstitial in the crystalline structure.

Our analysis in Fig. 4 extends the long-standing earlier assumption that amorphous materials contain structural building blocks of the corresponding crystalline phases.[1,61] We therefore suggest that amorphous phases can be thought of as containing building blocks of crystalline phases *and of defects* in crystals.

**From defects to defect clusters**

Although the majority of over-coordinated defects in a-Si exist as isolated $N = 5$ atoms, we find a stronger tendency for those atoms to occur close together than would be expected for a random placement of defects in a CRN. The energy distributions in Fig. 5a, now evaluated for defects and their local environments up to 4‴ sites, show that such clustering reduces the total average energy associated collectively with defect atoms and their immediate neighbourhoods. This is largely a result of clustered defects having fewer nearest neighbours per 5-fold defect. Neighbours of defects, on average, have an elevated energy compared to bulk a-Si. To avoid the extensivity of the excess energy with surface area when referred to the ground-state crystal energy, we reference the energies in Fig. 5a to the average value for bulk a-Si. In this way, the neighbours of a defect will only increase the energy of the defect cluster if they have an average energy above that of bulk a-Si. Hence, we can converge the prediction of the cluster excess energy, so that it becomes independent of the number of neighbours considered, by including neighbours that are sufficiently distant from the defect centre as to be typical of the bulk. In practice, this is achieved by including the 4‴ atoms. The widths of the distributions in Fig. 5a get narrower with cluster size because larger clusters have a greater total number of 4′, 4″ etc. neighbours (although fewer per 5-fold atom). The relationship between probability and cluster size is approximately exponential (Fig. 5b), with the exception of clusters of size 2, which are disfavoured compared to 3-membered clusters.

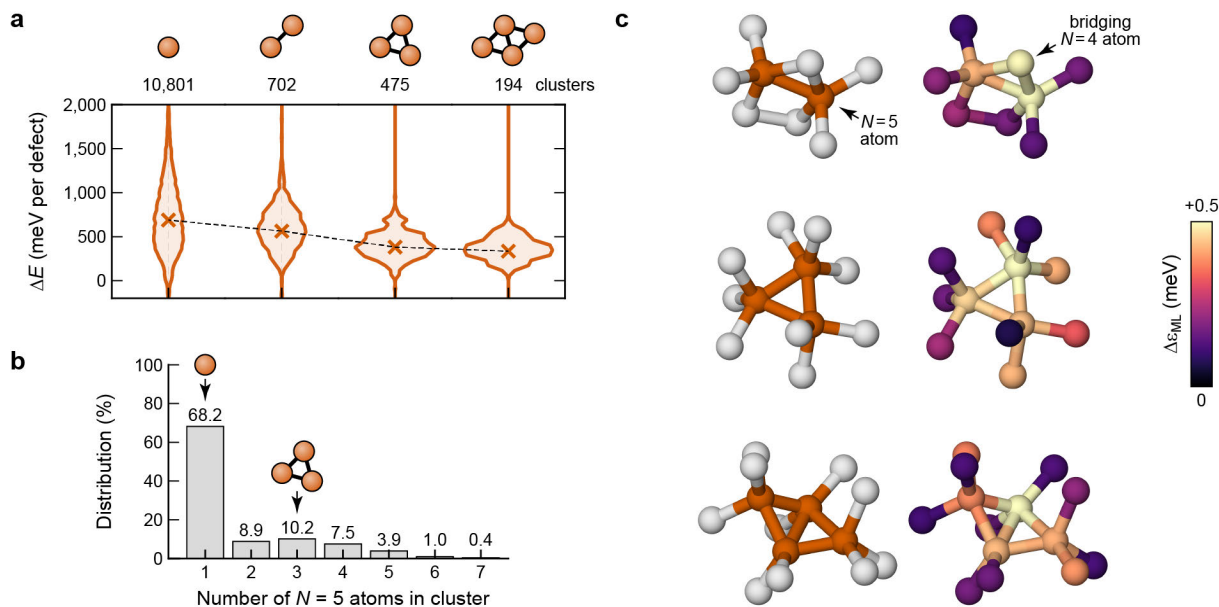

**Fig. 5: Clustering of defects in amorphous silicon.** (**a**) ML-predicted energy distributions for the most common 5-fold defect clusters, with the number of occurrences of each structure directly above each bar. Defect cluster energies are calculated by summing the individual atomic energies of defect cores and their immediate topological neighbours up to 3 bonds away relative to the mean a-Si energy and are reported per-coordination defect: viz. $\Delta E = \sum_i (E_i - \bar{E})/n_5$ where $i$ indicates the topological classification of atom $i$ and runs over 5, 4′, 4″, and 4‴; and $n_5$ is the number of 5-fold atoms in the cluster. (**b**) Statistics for the number of occurrences of clustered 5-coordination defects. (**c**) Examples of clustered $N = 5$ defects of different sizes, colour-coded by coordination numbers of the atoms (*left*) and by their ML local energy above diamond-type Si, $\Delta\varepsilon_{ML}$ (*right*). Note the high atomic energy of the bridging $N = 4$ atom in the top part of the panel, indicated by an arrow.

A possible qualitative explanation for this difference is provided by the examples in Fig. 5c: in the first case, a pair of $N = 5$ atoms is connected via a bridging $N = 4$ atom which in itself has a rather high atomic energy (highlighted by an arrow), consistent with the strain formed in the associated three-membered ring. In the larger fragments shown in Fig. 4c, there are still three-membered rings, but they are now formed by three and four $N = 5$ atoms clustering together, respectively. In these cases, the directly-connected $N = 4$ atoms have lower energies. The clustering of 5-coordinate defects can be interpreted as causing a reduction in the defective 'surface area'.

## Discussion

Our analyses support a comprehensive picture of defects in amorphous silicon. On one hand, 3-fold connected 'dangling-bond' defects are high in energy and do not strongly affect their surroundings. On the other hand, 5-fold connected defects are associated with a broad range of possible structures—which we can understand, at one extreme, as being similar to a trigonal bipyramid with an even distribution of bond distances, and at the other extreme, as being similar to the floating-bond description applied previously. 5-fold defects also have an extended influence on their surroundings, reaching beyond their immediate atomic neighbour environment, which explains their observed tendency to cluster together.

Beyond silicon, our study has more general implications for materials modelling. Defects and impurities in solids occur at a concentration of a couple of percent at most, yet they are highly consequential for the performance of functional materials. The structure and chemical bonding of these defects will likely differ markedly from the bulk, requiring a quantum-mechanically accurate and transferable description of the potential-energy surface associated with defect formation. In the present work, we have showcased how ML interatomic potentials can be used to study rare events in a systematic and statistical way, and how ML-predicted atomic energies can explain the stability of defect environments in one of the most widely studied amorphous materials, *viz.* a-Si. Qualitatively, this means a step away from idealised CRN models or small-scale simulation systems which (necessarily) contain only a handful of defects, moving towards a fully realistic description of the amorphous state.[62] In the future, we envisage similar ML-driven studies of defects and defect complexes that occur in a wide range of functional materials.

## Methods

**Teacher–student ML potentials.** We generate fast and robust interatomic potential models by using one ML model to 'teach' another. This idea is related to knowledge distillation in neural networks, albeit defined more generally. In ref. [29], we demonstrated it for a kernel-based method, *viz.* the Gaussian approximation potential (GAP) framework,[38] as the teacher model, and a linear fitting technique, *viz.* the moment tensor potential (MTP) approach,[63] as the student model. We used the indirectly-learned $M_{16}''$ model to generate structures via MD, and we use local energies predicted by the teacher GAP model[64] for analysis. Using the $M_{16}''$ local energies gave qualitatively similar results.

**Molecular dynamics.** MD simulations were carried out using the LAMMPS software,[65] interfaced to the GAP[38] and MTP[63] codes, respectively. The timestep was 1 fs. In ref. [29], the melt–quench simulations were conducted using the same variable-rate protocol as described in ref. [27], that is, at a quenching rate of $10^{11}$ K s$^{-1}$ for the most relevant parts of the temperature ramp, and $10^{13}$ K s$^{-1}$ elsewhere. The same type of analysis of defects performed herein on the million-atom structure cooled at $10^{11}$ K s$^{-1}$ (taken from ref. [29]) was performed on a separate structure produced by annealing a rapidly quenched ($10^{13}$ K s$^{-1}$) structure at 850 K. The results, which are equivalent to those in the main article, may be found in Supplementary Figs. 4–5.

**Electronic-structure analysis for validation.** We analysed with DFT the electronic structure of a small scale (512 atom) structural model made in the same way as the million-atom model to further validate structures produced with the ML potential. A clear energy gap between valence and conduction bands and exponentially shaped Urbach tails are indicators of high quality structural models of a-Si.[50] In our model, Urbach tails are evident and dangling bonds produce electronic states near the middle of the gap (Supplementary Fig. 3). Floating bonds in these models, which are thought to play a role in charge-carrier transport,[66] do not produce localised states in the gap, and as such would not be observed in an electron spin resonance experiment, similar to observations in ref. [67]. However, we observe that tail states are distributed among 4′ atoms in the vicinity of the floating bonds (Supplementary Fig. 3). This analysis suggests that only the dangling-bond sites would yield an electronic signature in the gap,[28] for which the concentration is low compared to other MD-derived structures at 0.7%.[19] Hence, the electronic properties of the million-atom model are consistent with other high-quality structural models. Further details of the electronic-structure calculations, including analysis of the ultra-large structure with a tight-binding Hamiltonian, are provided in the Supplementary Information. As an aside, we note that regions of strain defined by a 5-fold-coordinated atom and its immediate 4′ neighbourhood are rather more mobile than the atoms themselves, which allows the strained regions to diffuse through the structure at an appreciable rate even after vitrification has slowed atomic diffusion. The time-average of this motion could further contribute to the exponential Urbach tails.

**Structural analysis.** The SOAP kernel quantifies the similarity of two atomic environments on a scale from 0 (entirely dissimilar) to 1 (identical up to a defined cut-off radius).[46] Here, the structural similarity to crystalline phases was evaluated using the QUIP and GAP codes (https://github.com/libAtoms/QUIP). The SOAP kernel hyperparameters were: radial cut-off, $r_{cut}$ = 2.7 Å (noting that all nearest-neighbour distances in this analysis are scaled to 2.5 Å); neighbour-density smoothness, $\sigma_{at}$ = 0.3 Å; truncation of the density expansion, $n_{max}$ = 16 and $l_{max}$ = 8. Structural visualisations were created using OVITO.[68] A perplexity of 300 was used for the t-SNE dimensionality reduction in Fig. 2. Consistent results were achieved over the range 200–350 (highest tested) and using DBSCAN[69] for clustering in addition to bisecting $k$-means. Ring statistics for the 1M structure are provided in Supplementary Figure 7, for which we developed an implementation of the algorithm described in ref. [70] to handle the large system size. This code is available via GitHub (https://github.com/MorrowChem/julia_rings).

## Acknowledgements

J.D.M. acknowledges funding from the EPSRC Centre for Doctoral Training in Inorganic Chemistry for Future Manufacturing (OxICFM), EP/S023828/1. A.L.G. gratefully acknowledges financial support from the E.R.C. (Grant 788144). The authors acknowledge the use of the University of Oxford Advanced Research Computing (ARC) facility in carrying out this work (http://dx.doi.org/10.5281/zenodo.22558), as well as support from the John Fell Oxford University Press (OUP) Research Fund. We are grateful for computational support from the UK national high performance computing service, ARCHER2, for which access was obtained via the UKCP consortium and funded by EPSRC grant ref EP/X035891/1. D.A.D. and C.U. employed BRIDGES-2 at the Pittsburgh Supercomputing Center through allocation DMR190008P and PHY230007P from the Advanced Cyberinfrastructure Coordination Ecosystem, supported by the U. S. National Science Foundation.

Supplementary Information for

# Understanding defects in amorphous silicon with million-atom simulations and machine learning

Joe D. Morrow,[1] Chinonso Ugwumadu,[2] David A. Drabold,[2] Stephen R. Elliott,[3]
Andrew L. Goodwin[1,*] & Volker L. Deringer[1,*]

[1] *Inorganic Chemistry Laboratory, Department of Chemistry, University of Oxford, Oxford OX1 3QR, UK*

[2] *Department of Physics and Astronomy, Nanoscale and Quantum Phenomena Institute (NQPI), Ohio University, Athens, Ohio 45701, United States*

[3] *Physical and Theoretical Chemistry Laboratory, Department of Chemistry, University of Oxford, Oxford OX1 3QZ, UK*

\* andrew.goodwin@chem.ox.ac.uk; volker.deringer@chem.ox.ac.uk

# Supplementary Figures

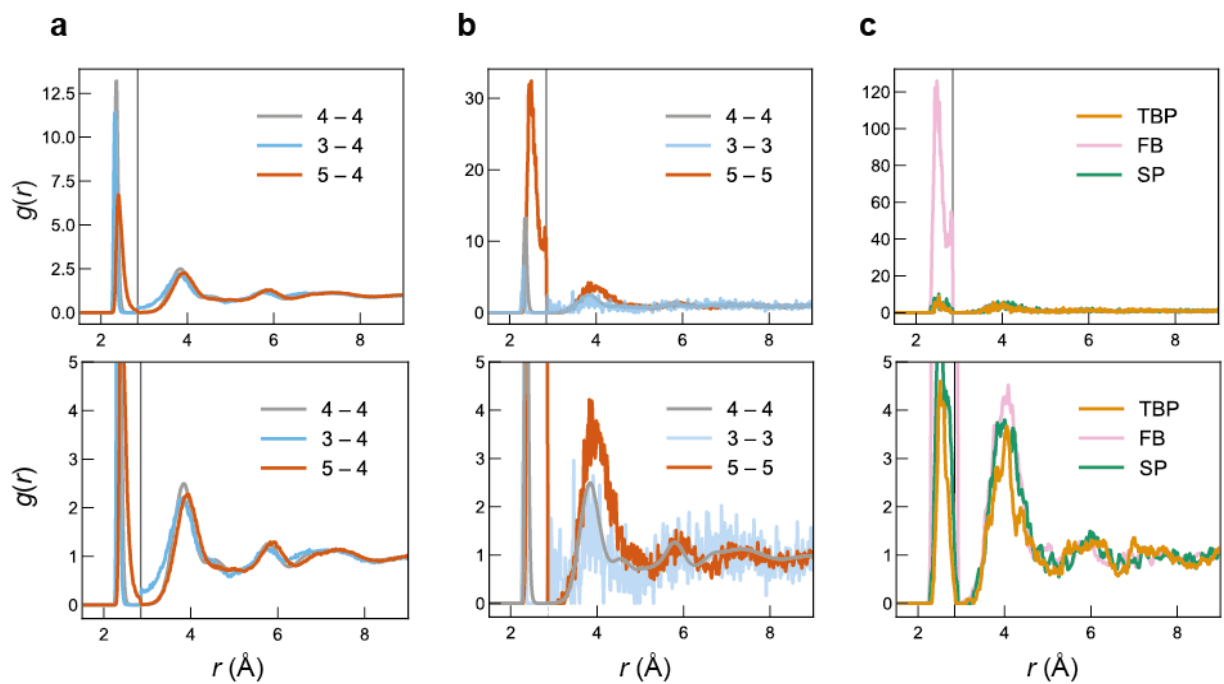

**Supplementary Fig. 1. Pair correlations for defect atoms.** In all columns, we show radial distribution functions, $g(r)$, in full (upper panels) and magnified (lower panels). (**a**) Correlations between each of 3-, 4-, and 5-fold connected atoms and 4-fold connected atoms. (**b**) Correlations between atoms of the same connectivity. (**c**) Correlations between 5-fold connected defects, separated by classification into trigonal-bipyramid-like (TBP), 'floating bond'-like (FB), and square-pyramidal-like (SP) as defined in Fig. 2 of the main text. Vertical lines indicate the radial cut-off used to define coordination numbers.

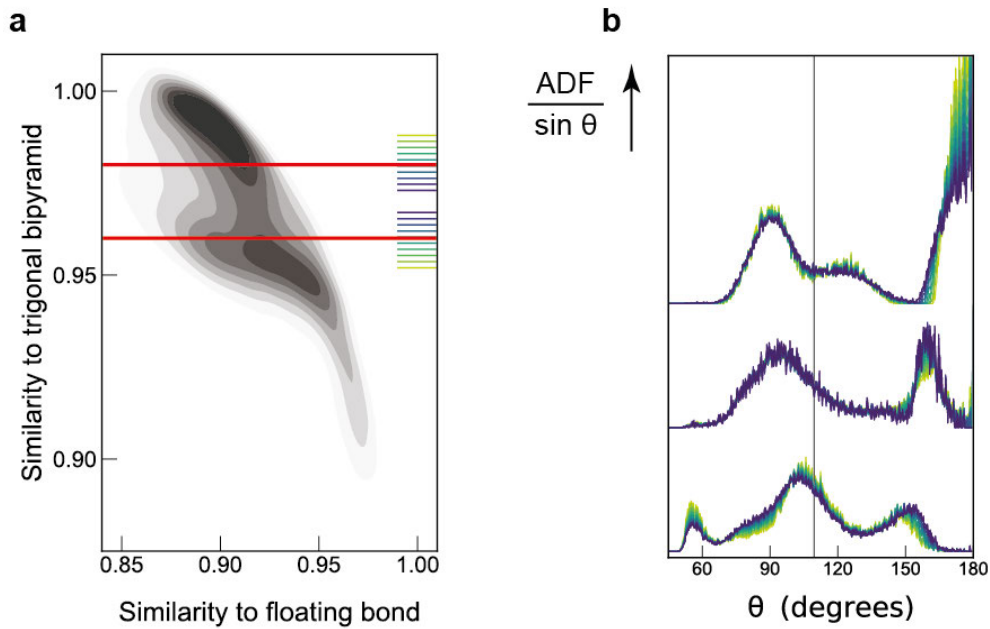

**Supplementary Fig. 2. Classification methods for defects.** (**a**) SOAP similarity values used for defining the three classes of fivefold defect (red lines) in Fig. 2 of the main text. Structures with a similarity greater than the upper red line are categorised as trigonal-bipyramid-like, those between the lines as square-pyramidal-like, and those below as 'floating bond'-like. (**b**) Sensitivity of ADFs to similarity cut-off values. ADFs are shown in colours purple to yellow corresponding to a range of similarity values in panel a of the same colour. The form of the ADF is quite insensitive over a wide range of similarity values; therefore the results in Fig. 2 are not strongly affected by the exact cut-off value for the similarity to distinguish classes.

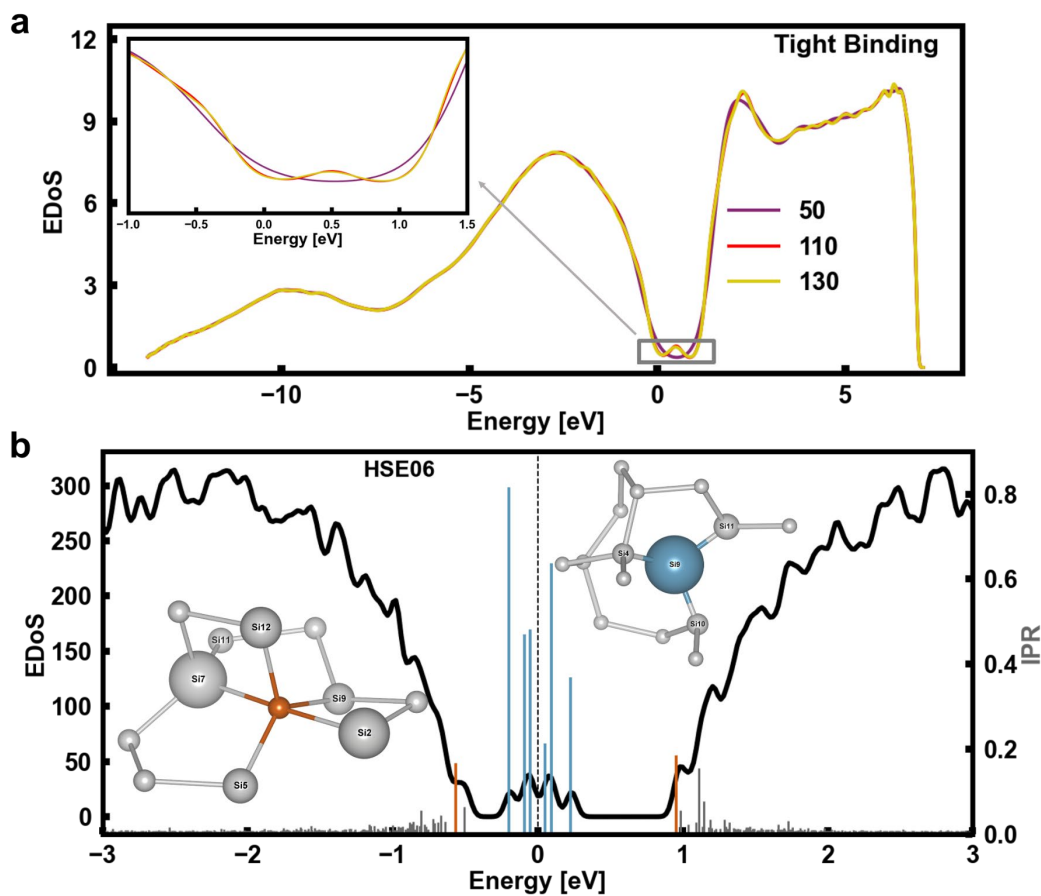

**Supplementary Fig. 3. Electronic structure of a-Si models.** (**a**) Electronic density of states (EDoS) for the 1-million atom a-Si model discussed in the main text, computed using the tight-binding methodology of ref. S1. The colouring of the lines refers to the number of moments used in the maximum-entropy reconstruction. See ref. S2 for a summary of the methodology and an example of a DOS computation on a similarly large (albeit crystalline) Si structure. (**b**) EDoS for a 512-atom a-Si model computed with the HSE06 hybrid functional.[S3–S4] The 512-atom model is constructed using the same protocol as for the 1-million atom model, and the two models exhibit gaps of similar width. The Fermi level is shifted to $E = 0$ eV in panel **b**. Localised states within the energy-gap region were obtained for the 512-atom model (blue and orange). States near mid-gap are localised exclusively on dangling bonds (blue spikes), consistent with earlier electronic-structure computations in ref. S5, while tail states are distributed among 4-fold atoms that are directly connected to floating bonds (orange spikes). The sizes of the spheres centred on atoms indicate the extent of localisation. The dangling-bond state with smallest localisation (just above the Fermi level) is somewhat delocalised by resonant mixing with 4-fold atoms directly connected to a floating bond, as discussed elsewhere.[S6]

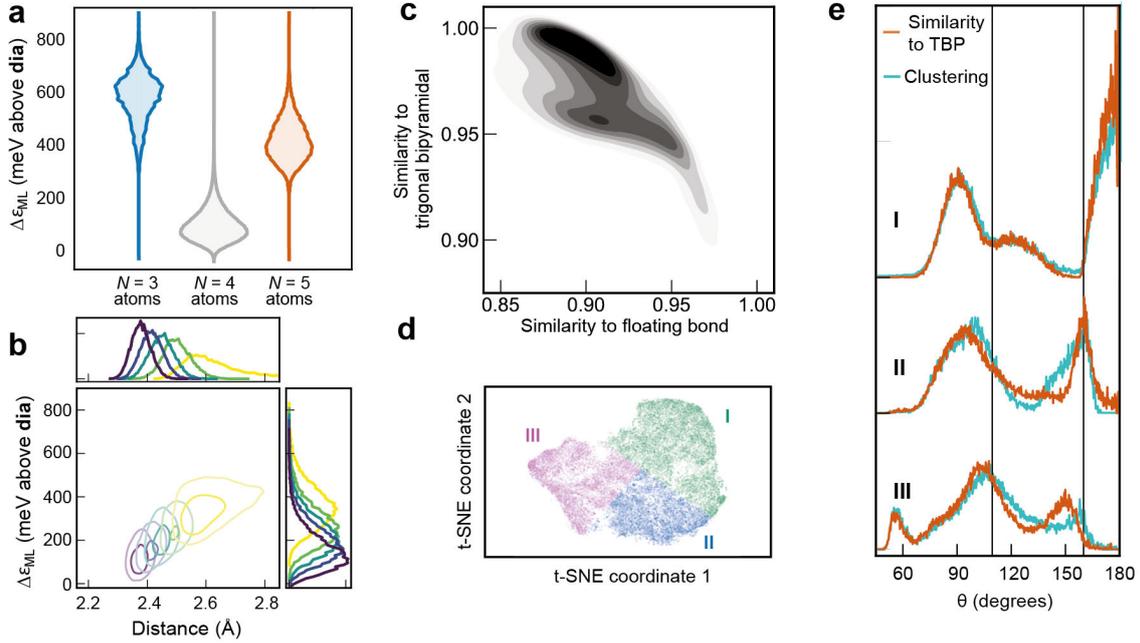

**Supplementary Fig. 4. An alternative 1-million atom structure from rapid quenching and annealing.** We repeat analyses as in the main text for a separate, entirely uncorrelated 1M atom structure, which was produced with a slightly different protocol: rapid cooling at $10^{13}$ Ks$^{-1}$ followed by a long annealing period of 2 ns at 840 K. This structure has a similar number of 5-fold atoms as the structure discussed in the main text, which was derived from slower cooling at $10^{11}$ K s$^{-1}$. (**a**) Distributions of the ML atomic energies, $\varepsilon_{ML}$, of defects in the model, shown separately for different nearest-neighbour coordination numbers, $N$. (**b**) Two-dimensional correlation plot of the neighbour density for 5-coordinated defects (as in panel **e**). (**c**) 2D plot of SOAP kernel similarity of $N = 5$ atoms in the a-Si model to the idealised TBP and floating-bond environments, respectively. The distribution of values for individual atoms is shown as a heat map. (**d**) Unsupervised classification of 5-fold atoms. The full distance matrix, $\mathbf{D} = \sqrt{2 - 2\mathbf{K}}$, is embedded in 2D with the dimensionality-reduction algorithm t-SNE,[S7] where $\mathbf{K}$ is the kernel matrix built from the similarity of each 5-fold atom with every other 5-fold atom. Bisecting $k$-means is used to identify clusters **I–III**. (**e**) Bond-angle distribution function (BADF) plots, scaled by $\sin\theta$, for atomic triples centred on all 5-fold coordinated atoms. The BADFs are plotted separately for the three distinct categories of 5-fold defects related to idealised structures respectively from top to bottom (as illustrated with selected examples of such configurations from the a-Si model). The 5-fold atoms are separated into these categories via two methods but with similar results: by comparison to the idealised structures of panels **a** and **b** via SOAP similarity (orange) and from unsupervised clustering (cyan). Vertical black lines at the ideal tetrahedral angle (109.5°) and at 160° are guides for the eye.

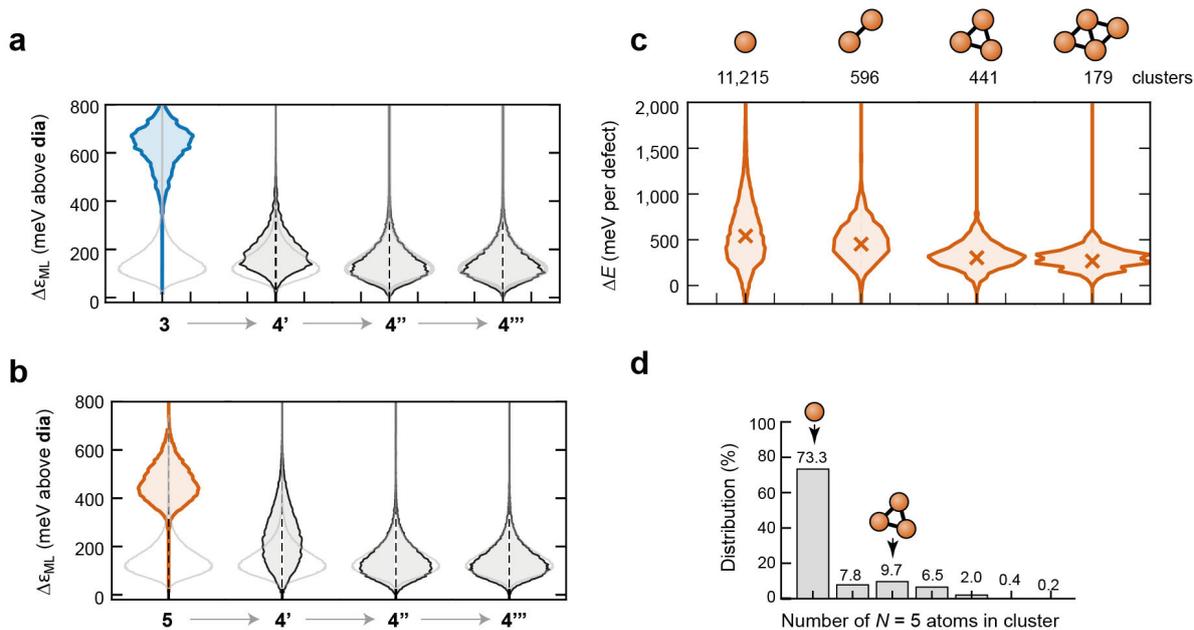

**Supplementary Fig. 5. Locality effects in the alternative 1-million atom structure.** Further analysis of the locality of defects for the annealed 1-million atom structure of Supplementary Fig. 4. (**a**) Distribution of ML local energies for $N = 3$ atoms and their surroundings. The distribution for bulk a-Si is shown in light grey. (**b**) Same but for $N = 5$ atoms and their surroundings. (**c**) ML-predicted energy distributions for the most common 5-fold defect clusters, with the number of occurrences of each structure directly above each bar. Defect cluster energies are calculated by summing the individual atomic energies of defect cores and their immediate topological neighbours up to 3 bonds away relative to the mean a-Si energy and are reported per-coordination defect: viz. $\Delta E = \sum_i (E_i - \bar{E})/n_5$ where $i$ indicates the topological classification of atom $i$ and runs over 5, 4′, 4″, and 4‴; and $n_5$ is the number of 5-fold atoms in the cluster. (**d**) Statistics for the number of occurrences of clustered 5-coordination defects.

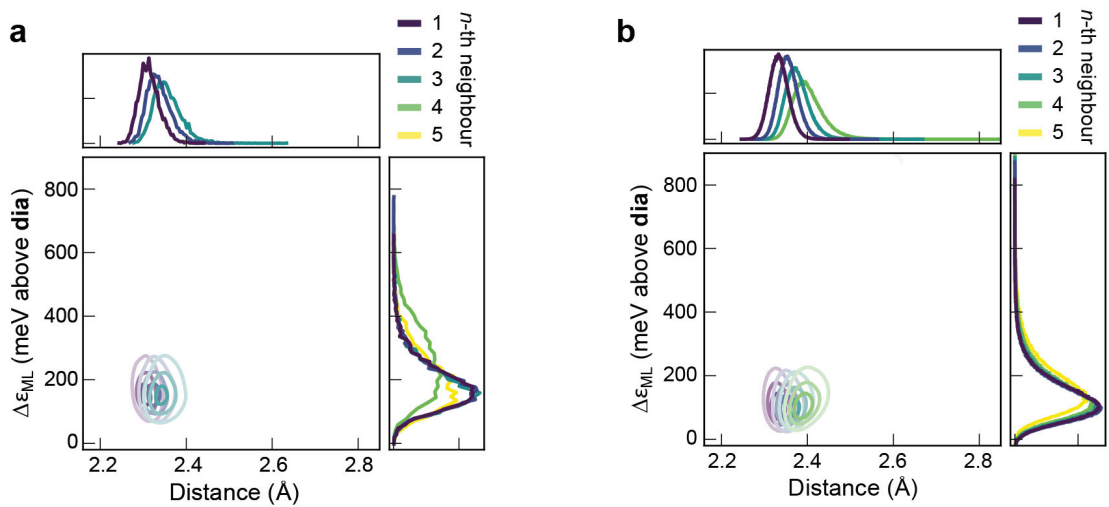

**Supplementary Fig. 6. Correlation plots for 3-fold and 4-fold coordinated atoms.** Two-dimensional correlation plot of the neighbour density for (**a**) 3-fold defects and (**b**) 4-fold co-ordinated atoms versus energy, given separately for the immediate neighbours ($n$ = 1–5, purple to green to yellow). Note that the 4-th and 5-th neighbours are off the *x*-axis scale for 3-fold connected atoms. Likewise for the 5-th neighbours of the 4-fold connected atoms. The energies of 5-th neighbours for 3-fold and 4-fold atoms are like that of bulk a-Si, which indicates that these more distant atoms are typical of the bulk, in contrast to the 5-th neighbours of 5-fold defects.

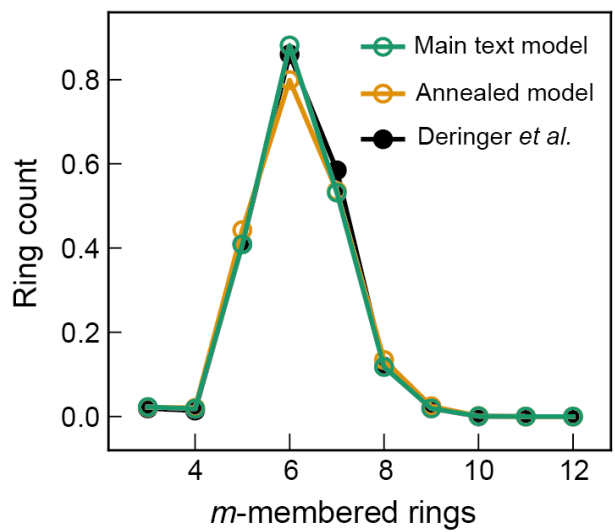

**Supplementary Fig. 7. Ring statistics for 1M atom models.** Ring counts per atom are displayed for both 1M atom structural models: the one discussed in the main text (green) and the alternative, newly created annealed structure introduced in Supplementary Fig. 4 (orange). The results are compared to those for a 4,096-atom reference structure (black) from Deringer *et al.* (ref. S8; obtained with the GAP-18 potential[S9]). The structure discussed in the main text was prepared using the same MD protocol as the 4,096-atom reference and they are here shown to have very similar ring statistics, providing further validation of the quality of the 1M atom model. The algorithm used to obtain ring counts is described in ref. S10 (see Methods section).

## Supplementary References